\documentclass[
amsmath,amssymb,
 aps,twocolumn, showpacs,print
]{revtex4-1}
\usepackage{graphicx}
\usepackage{dcolumn}
\usepackage{bm}
\usepackage{hyperref,epstopdf}
\usepackage{float}
\begin{document}
\title{\large Nuclear shape phase transitions within a correlation between two quantum concepts  }
\author{S. Ait Elkorchi}
\author{M. Chabab}
\author{A. El Batoul}
\author{A. Lahbas}
\author{M. Oulne}
\email{Oulne@uca.ac.ma}
\affiliation{High Energy Physics and Astrophysics Laboratory, Department of Physics, Faculty of Sciences Semlalia, Cadi Ayyad University P.O.B 2390, Marrakesh 40000, Morocco}
\date{\today}
\begin{abstract}
We present a correlation that we have revealed, for the first time, between both quantum concepts, namely: the Minimal Length (ML) and the Deformation Dependent Mass (DDM) in transitional nuclei near the critical points symmetries (CPS) X(3) and Z(4). Such a correlation could be considered as a new signature for these CPS. This new signature allowed us to predict new candidate nuclei for these critical points.
\end{abstract}
\pacs{21.60.Re, 21.60.Ev, 21.60.Fw, 27.60.+j, 27.70.+q, 27.80+w, 27.90.+b}
 \keywords{Bohr–Mottelson model, Critical point symmetries, Minimal length, Davidson potential}
\maketitle
\section{Introduction}
Nowadays, the shape phase transitions in nuclei occupy a salient place in nuclear structure research and continue to attract interest of scientists. They have been first introduced within the interacting Boson Model (IBM) \cite{b1}. In its original version, this model allows describing  collective excited states of nuclei in terms of bosons of angular momentum $L=0$ (s-boson) and  $L=2$ (d-boson) in the context of the parent symmetry group U(6), which has limiting dynamical symmetries, namely: U(5) for the vibrator collective motion mode, SU(3) for axial rotor and O(6) for $\gamma$-unstable case. However, most nuclei deviate from these ideal limits and sit between them. Such a fact paved the way for studying transitions from one phase to another. Therefore, critical points symmetries (CPS) for these shape phase transitions have been born \cite{b2}. The pioneering ones amid them were E(5) \cite{b3} and X(5) \cite{b4} corresponding to the shape phase transitions U(5)$\leftrightarrow$O(6) and U(5)$\leftrightarrow$ SU(3) respectively. Later, a $ \gamma $-rigid (with $ \gamma =0 $) version of X(5), called X(3), has been introduced \cite{b5}. In the same way, other CPS have been developed like for example Z(5) and its $\gamma$-rigid version Z(4) (with $ \gamma =\pi/6 $) corresponding to shape phase transitions from prolate to triaxial symmetry \cite{b6}. The above-mentioned dynamical symmetries are located at the vertices of the so-called Casten triangle, while the most of CPS are on its sides and are parameter-free solutions, where the square well potential has been used. For such a peculiar feature, they are considered as verifiable points for the experiment. Based on these CPS, several theoretical studies have been carried out within IBM \cite{b1,b6a} or Bohr Mottelson model \cite{b7}. In the framework of this latter, many additional attempts have been devoted to obtain solutions of the Bohr Hamiltonian with a constant mass \cite{b8a1,b8a2,b8a3,b8a4} as well as within the DDM concept \cite{b8b1,b8b2,b8b3,b8b4,b8b5}. Such a concept \cite{b9}, which is widely used in quantum physics, is equivalent to a deformation of the canonical commutation relations:
\begin{equation}
[x_i, x_j] = 0,\ 
[x_i, p_j] =  {\rm i} \hbar \delta_{i,j},\ 
[p_i, p_j] =  0 \,  
\label{eq1}
\end{equation}
where $i=1,2,3$. By replacing the momentum components
$p_i = - {\rm i} \hbar \nabla_i= - {\rm i} \hbar \partial/\partial x_i$  by some deformed hermitian operators:
\begin{equation}
\pi_i = \sqrt{f(x)}\, p_i \sqrt{f(x)}\,  
\label{eq2}
\end{equation}
where the positive real deforming function $f(x)$ depends on the coordinates $ x=(x_{1} ,x_{2},x_{3})$, both last commutators in Eq.\eqref{eq1}  transforme into :
\begin{eqnarray}
[x_i, \pi_j]  = {\rm i} \hbar f(x) \delta_{i,j}, \
[\pi_i, \pi_j]  = {\rm i} \hbar [f_j(x) \pi_i-f_i(x) \pi_j]
\label{eq3}
\end{eqnarray}
with $f_i(x) \equiv \nabla_i f(x)$.\\
Recently, a great interest has been consecrated to the quantum mechanical problems related to a generalized modified commutation relations involving a minimal length or generalized uncertainty principle \cite{b10,b11,b12,b13,b13aa}:
\begin{equation} 
\Delta X\Delta P\geq\dfrac{\hbar}{2}(1+\alpha(\Delta P)^{2})\, 
\label{eq4}
\end{equation}
obtained from the deformed canonical commutation relation:
\begin{equation}
[X,P]=i\hbar(1+\alpha P^{2})\,
\label{eq5}
\end{equation}
where $\alpha $ represents the minimal length parameter($\alpha<1 $). In our pioneering work \cite{b13}, we have introduced this concept, for the first time, in nuclear structure through Bohr-Mottelson model. Now, since both above-presented concepts are connected to deformed canonical commutation rules or curved space, it appears to be legitimate to ask the question : \textbf{does exist a correlation between them ?} The answer to this question is the purpose of the present paper in which we will show the existence of such a correlation through solutions of Bohr-hamiltonian for transitional nuclei in the limits of CPS X(3) and Z(4). The revealed correlation will be exploited as a new signature for prediction of new candidate nuclei to these symmetries.
\section{Method and Results }
\begin{figure}[h]
	\begin{center}
		\includegraphics[width=70mm,height=40mm]{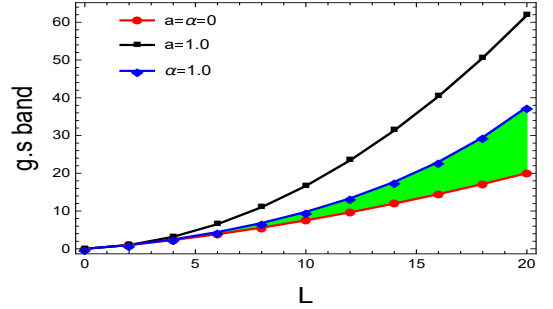}
		\caption{Ground state band energy of $X(3)$, normalized to the first excited level given as a function of angular momentum L  for the parameters $\alpha$ and $a$ ranging from 0 to 1. The colored area represents an overlap between the X(3)-ML and X(3)-DDM models.}
		\label{fig1}
	\end{center}
\end{figure}

\begin{figure}[!h]
	\begin{center}
		\includegraphics[height=40mm]{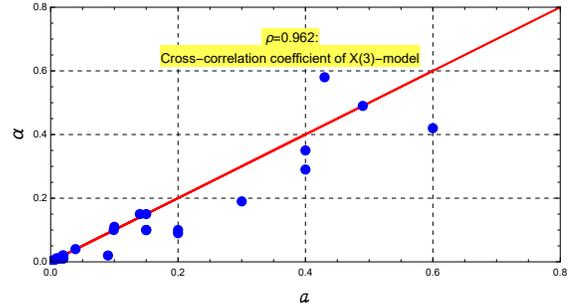}
		\caption{The Correlation between the parametres $a$ and $\alpha $ in X(3).}
		\label{fig2}
	\end{center}
\end{figure}

\begin{figure}[h]
	\begin{center}
		\includegraphics[height=40mm]{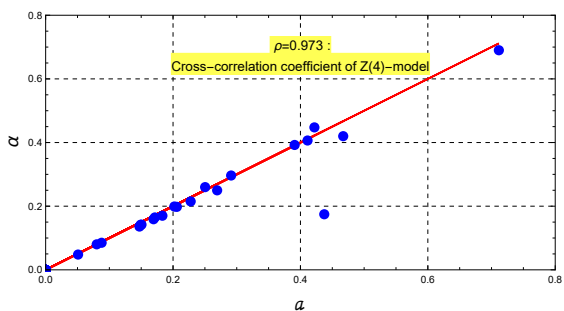}
		\caption{The correlation between the parametres $a$ and $\alpha $ in Z(4).}
		\label{fig3}
	\end{center}
\end{figure}

\begin{figure}[!h]
	\begin{center}
		\includegraphics[height=40mm]{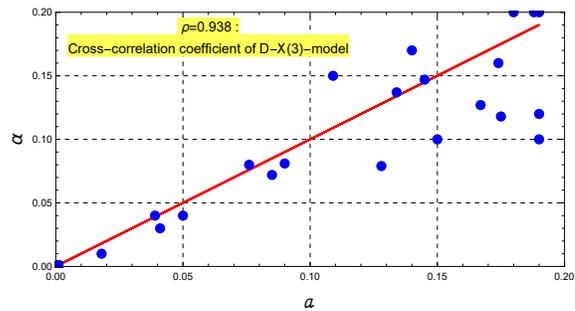}
		\caption{The correlation between the parametres $a$ and $\alpha $ in D-X(3).}
		\label{fig4}
	\end{center}
\end{figure}

\begin{figure}[!h]
	\begin{center}
		\includegraphics[height=40mm]{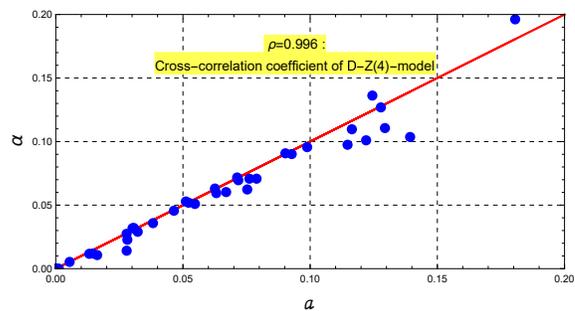}
		\caption{The correlation between the parametres $a$ and $\alpha $ in D-Z(4).}
		\label{fig5}
	\end{center}
\end{figure}

\begin{figure}[h]
	\begin{center}
		\includegraphics[height=40mm]{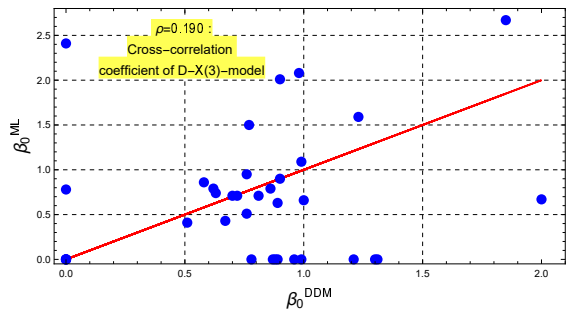}
		\caption{The correlation between Davidson potential's minimum $\beta_{0}^{\rm{ML}}$ and $\beta_{0}^{\rm{DDM}}$ in X(3) symmetry. }
		\label{fig6}
	\end{center}
\end{figure}

\begin{figure}[!h]
	\begin{center}
		\includegraphics[height=40mm]{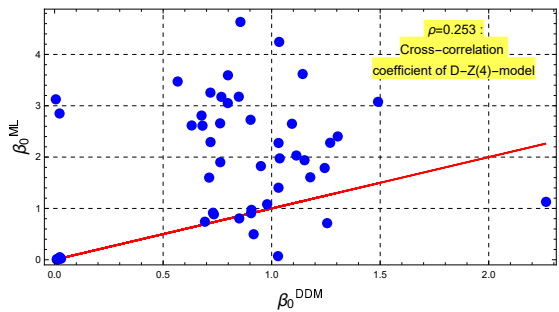}
		\caption{The correlation between Davidson potential's minimum $\beta_{0}^{\rm{ML}}$ and $\beta_{0}^{\rm{DDM}}$ in Z(4) symmetry.}
		\label{fig7}
	\end{center}
\end{figure}

The original Bohr Hamiltonian \cite{b7} is written as :
\begin{equation}
\begin{split}
& H = -\frac{1}{2B}\left[ 
{1\over \beta^4} {\partial \over \partial \beta} 
\beta^4 {\partial \over \partial \beta} 
+{1\over \beta^2 \sin 3\gamma} {\partial \over \partial \gamma} 
\sin 3\gamma {\partial \over \partial \gamma} \right.  \\
& \left. -{1\over 4 \beta^2} 
\sum_{k=1,2,3} {L_k^2 \over \sin^2\left(\gamma -{2\over 3} \pi k \right)}
\right]+v(\beta,\gamma),
\end{split}  
\label{eq6a}
\end{equation}
where $\beta$ and $\gamma$ are the deformation variables, $B$ is the collective mass parameter, and $L_k(k =1, 2, 3)$ are the projections of the angular momentum on the body-fixed k-axis. This famous Hamiltonian describes a variety of  different types of collective motion depending on the potential energy function $v(\beta,\gamma)$ and the inertial parameters. On the other hand, the fundamental equations of Bohr Hamiltonian (\eqref{eq6a}) within both concepts and their solutions  are already established in several papers \cite{b8b1,b8b2,b8b3,b8b4,b8b5,b13,b13a}. The deforming function $f(\beta)$, which depends on the form of the used potential, is chosen in the case of an infinite square well with a null depth in the following form:
\begin{equation}
f(\beta)=\beta^{-a},\   a\in\left[0,1\right],
\label{eq6}
\end{equation}
where $a$ is a DDM parameter. This parameter is not a simple extra additional one for fitting experimental data, but it is a model's structural parameter as it has been shown in \cite{b8b4,b8b5}. An important consequence of this chose is the possibility to derive the energies spectrum, as function of zeros of the Bessel functions, employing the same method in Refs\cite{b5,b6}. Thus, the energies of X(3) model, characterized by the principal quantum number $s$ together with total angular momentum $L$ are given in this frame by:
\begin{equation}
E_{s,L} = \frac{\hbar^2}{2B_m}\bar{k}_{s,\eta}^2,\ \ \bar{k}_{s,\eta}= \frac{\chi_{s,\eta}}{\beta_{\omega}^{a+1}} \ ,
\label{eq7}
\end{equation}
where $\chi_{s,\eta}$ is the $s$-th zero of the Bessel function of the first kind
$J_{\eta}(\bar{k}_{s,\eta}\beta_{\omega})$ and $\beta_{\omega}$ is the potential's width. $\eta$	is a parameter given by :
\begin{equation}
\eta=\frac{\sqrt{a(a+1)+\frac{L(L+1)}{3}+\frac{1}{4}}}{2(a+1)}
\label{eq8}
\end{equation} 
This a new formula, which is not found in the literature.
In the ML concept, the eigenenergies formula reads \cite{b13}:  
\begin{equation}
E_{s,L} = \frac{\hbar^2}{2B_m}\times\frac{\bar{k}_{s,\eta}^2}{1-2\hbar^2\alpha\bar{k}_{s,\eta}^2},\ \ \bar{k}_{s,\eta}= \frac{\chi_{s,\eta}}{\beta_{\omega}}
\label{eq9}
\end{equation}
where the parameter $\eta$, in this case, is given by :
\begin{equation}
\eta=\left(\frac{L(L+1)}{3}+\frac{1}{4}\right)^{\frac{1}{2}}.
\label{eq10}
\end{equation}
If preliminarily, we plot in the same figure (Fig.\ref{fig1}) the eigenenergies given in Eq.\eqref{eq7} and Eq.\eqref{eq9} versus the angular momentum for arbitrary values of the parameters $a$ and $ \alpha $ and for $\beta_{\omega}$=40, one can see that the allowed regions for both concepts (ML and DDM) exhibit an overlap which is located in a region close to the CPS X(3). This observation still incites the curiosity about the searched correlation between them.
Now, dealing with concrete nuclei, we have calculated the energy ratios $ R_{L_{g}/2_{g}} $ and $ R_{L_{\beta}/2_{g}} $ of different levels $ L_{g} $ and $ L_{\beta} $ of the ground state (g.s) and $ \beta $ bands, respectively, normalized to the first excited level of the g.s band, for 36 even-even nuclei belonging to the following isotope chains:  $^{104}$Ru, $^{106}$Cd, $^{112}$Pd, $^{106,116-120}$Cd, $^{116-134}$Xe, $ ^{132,138}$Ce, $^{132-136,142}$Ba,  $ ^{140-144}$Gd,  $ ^{152}$Gd, $ ^{154}$Dy, $ ^{172}$Hf,  $^{172,176}$Os, $ ^{190}$Os, $^{186-190}$Pt, $ ^{194-196}$Pt, $ ^{140,148}$Nd.
We have chosen the nuclei for which the ratio $ R_{4g/_{2g}} $ is not far from the value 2.44, which is a reference point for the X(3) model. The obtained values for the parameters $a$ and $ \alpha $ from the fit on all available experimental data are depicted  in Fig.\ref{fig2}.
Indeed, this figure shows a strong correlation between ML and DDM concepts where the cross-correlation coefficient is equal to 0.96. The nuclei situated on or close to the bisectrix have been proved to be the best candidates for X(3) symmetry \cite{b5,b8a4} namely $ ^{172}$Os, $^{120,126}$Xe, $^{148}$ Nd and $^{186}$Pt, including two new ones : $^{124}$Xe and $ ^{190}$Pt . For these nuclei, particularly, the parameters $a$ and  $ \alpha $  are nearby 0.001. This value corresponds to the lower boundary of the allowed region for both concepts (Fig. \ref{fig1}) under which their effect is canceled.
\\ As to the CPS Z(4) within DDM concept, the eigenvalues are determined by the following formula:
\begin{equation}
E_{s,L} = \frac{\hbar^2}{2B_m}\bar{k}_{s,\eta}^2,\ \ \bar{k}_{s,\eta}= \frac{\chi_{s,\eta}}{\beta_{\omega}^{a+1}} \ ,
\label{eq11}
\end{equation}
with,
\begin{equation}
\eta=\frac{\sqrt{L(L+4)+3n_{\omega}\left(2L-n_{\omega}\right)+2a(3a-4)+4}}{2(a+1)}.
\label{E12}
\end{equation}
where $n_w$ is the wobbling quantum number, 
while in the ML concept, the equations above are defined respectively by \cite{b13}:
\begin{equation}
E_{s,L} = \frac{\hbar^2}{2B_m}\times\frac{\bar{k}_{s,\eta}^2}{1-2\hbar^2\alpha\bar{k}_{s,\eta}^2},\ \ \bar{k}_{s,\eta}= \frac{\chi_{s,\eta}}{\beta_{\omega}}
\label{eq13}
\end{equation}
and,
\begin{equation}
\eta=\frac{\sqrt{L(L+4)+3n_{\omega}\left(2L-n_{\omega}\right)+4}}{2}.
\label{eq14}
\end{equation}
The calculations of the energy ratios for all bands, by the above equations  have been carried out for several isotopes, namely: ${}^{98-104}$Ru, ${}^{102-116}$Pd, ${}^{106-120}$Cd, ${}^{118-134}$Xe, ${}^{130-136}$Ba, ${}^{134-138}$Ce, ${}^{142}$Ba, ${}^{142-144}$Gd, ${}^{152}$Gd, ${}^{186-200}$Pt, for which the ratio  $ R_{4_{g}/2_{g}} $ is nearby 2.23. This value is a reference point for the Z(4) model. The obtained paramters a (DDM) and $ \alpha $ (ML), by fitting Eq.\eqref{eq11} and Eq.\eqref{eq13} on all available experimental levels, are plotted in Fig.\ref{fig3}.
\\ Here again, one can observe a strong correlation between both concepts (DDM and ML). The cross correlation coefficient is equal to 0.97. The best candidate nuclei for this CPS are set on or close to the bisectrix, which are: ${}^{128-132}$Xe  and ${}^{192-196}$Pt. These isotopes have been already proved to be the best candidates for Z(4) model \cite{b8b5,b8a3} including the new one $^{114}$Pd. However, we have to notice that even if the eventual existence of $ \gamma $-rigid triaxial nuclei in nature remains a hypothetical issue, the aim of their treatment  in the present study is to check the above found correlation  between ML and DDM concepts. The obtained values of the parameters $a$ and $ \alpha $ are also nearby 0.001 like in X(3) symmetry. Here, we have to bear in mind that Eq.\eqref{eq9} and Eq.\eqref{eq13}, normalized to the first excited level of the g.s. band, were fitted for both parameters $ \alpha $ and $\beta_{w}$, while in Eq.\eqref{eq7} and Eq.\eqref{eq11} the parameter  $\beta_{w}$ is simplified in the energy ratios $R_{L_{g}/2_{g}} $, $R_{L_{\beta}/2_{g}}$ and $R_{L_{\gamma}/2_{g}}$. However, the correlation between the concepts parameters $a$ and $ \alpha $ was not negatively influenced  by the potential parameter $\beta_{w}$.  Therefore, in order to see further whether the above found correlation is or not negatively impacted by the form or type of the used potential, we apply the above concepts to Davidson potential \cite{b13a}. It should be noted however that the treatments of all calculations for both concepts ( DMM and ML) for this potential are similar to those presented in references \cite{b8b5} and \cite{b13a} respectively.
The fit of the corresponding formulas for the energy ratios, in the two models X(3) and Z(4), on the available experimental data \cite{b18} for all above used isotope chains has lead to the parameters values of $a$ and $\alpha $, which are depicted in Fig.\ref{fig4} and Fig.\ref{fig5}. 
From these figures, it is obvious that a very strong correlation exists between the two quantum concepts. The cross-correlation coefficient is 0.97 in the X(3) case and 0.99 in the Z(4) one. The best candidate nuclei for both models are the same as the above cited ones and also lay on or close to the bisectrix. The corresponding parameters $a$ and $ \alpha $ are this once spread between 0.07 and 0.2. It is worthful to point out that many of the obtained values of the parameters $a$ and $ \alpha $ are very close to each other, so the corresponding points on correlation figures overcome. Moreover, in Fig.\ref{fig6} and  Fig.\ref{fig7}, we presented the Davidson potential's minimum $ \beta_{0} $(ML), which is obtained within ML concept, versus $ \beta_{0} $ (DDM) corresponding to the DDM one. 
\\
From these figures, one can see that there is a weak correlation between both minima. The cross-correlation coefficient is less than 0.3. Thus, this result corroborates the fact that the revealed correlation between both quantum concepts (ML and DDM) is not destructively affected by the presence of other model's parameters and hence its existence is independent of the form or type of the used potential.
\section{SUMMARY}
 The study in the present paper has confirmed that the two  quantum concepts, namely: the ML and DDM, which are related to deformed commutation relations or space curvature, are well and truly strongly correlated. The uncovered correlation has been used as a new signature for some nuclear CPS allowing us to predict new candidate nuclei to these symmetries. The present revelation will pave the way for further investigations of such correlation in other shape phase transitions in nuclei at other CPS. 

\end{document}